\documentclass[preprint, 10pt, 5p]{elsarticle}
\usepackage{geometry} 
\usepackage{graphicx} 
\usepackage{amsmath}
\usepackage{amssymb}
\usepackage{float} 
\usepackage{wrapfig} 
\usepackage{lipsum} 
\usepackage[hidelinks]{hyperref}

\graphicspath{{figures/}} 

\newcommand{\bv}{\mathbf{v}}

\newcommand{\br}{\mathbf{r}}

\newcommand{\bB}{\mathbf{B}}

\newcommand{\D}{\text{d}}

\begin{document}
\title{Temperature Screening and Cross-Field Impurity Accumulation from a Thermodynamic Perspective}
\date{\today}
\author{E.~J.~Kolmes}
\ead{ekolmes@princeton.edu}
\address{Department of Astrophysical Sciences, Princeton University, Princeton, New Jersey 08544, USA}
\author{I.~E.~Ochs}
\address{Department of Astrophysical Sciences, Princeton University, Princeton, New Jersey 08544, USA}
\author{M.~E.~Mlodik}
\address{Department of Astrophysical Sciences, Princeton University, Princeton, New Jersey 08544, USA}
\author{N.~J.~Fisch}
\address{Department of Astrophysical Sciences, Princeton University, Princeton, New Jersey 08544, USA}

\begin{abstract}
In a variety of different systems, high-$Z$ ion species show a marked tendency to accumulate in regions of high plasma density. It has previously been suggested that the apparent universality of this behavior could be explained thermodynamically, in terms of the maximum-entropy state attainable when the system must obey an ambipolarity condition. However, the previous treatment did not allow for the possibility of temperature gradients. 
Here, tools from non-equilibrium (Onsager) thermodynamics are used to show that ambipolarity continues to play a key role in producing this behavior in the presence of temperature gradients, and to recover well-known temperature screening effects that appear in these cases. 
\end{abstract}


\maketitle

\section{Introduction}

The relative accumulation of different ion species is a topic of interest -- and, often, of serious concern -- for a variety of plasma technologies. In fusion devices, it is important to ensure that fuel ions are mixed in the high-temperature region of the plasma. At the same time, it is necessary in steady-state operation to flush out fusion products (ash) lest they choke out the reaction. Meanwhile, the excessive accumulation of high-charge-state impurities can result in large radiative energy losses, and must be avoided. These issues are similarly important for many non-fusion applications. For instance, for plasma mass filters (which are designed to separate out different constituents of a plasma based on mass), control over the relative transport of different particles is the most essential feature of the device. 

In isothermal magnetically confined plasmas, in regimes where turbulence does not dominate, there is a classic result \cite{Spitzer1952, Taylor1961ii, Braginskii1965} that in steady state the density profiles of two ion species satisfy 
\begin{gather}
n_a^{1/Z_a} \propto n_b^{1/Z_b} . \label{eqn:pinch}
\end{gather}
Here $n_a$ is the density profile of species $a$, $n_b$ is the density profile of species $b$, and $Z_a e$ and $Z_b e$ are the two species' charge states (where $e$ is the elementary charge). 
The densities are assumed to vary only in the direction perpendicular to the magnetic field. 
Eq.~(\ref{eqn:pinch}) predicts a dramatic accumulation of high-$Z$ species in the highest-density regions of a plasma. 
In the presence of some potentials $\Phi_a$ and $\Phi_b$ affecting the two species (for example, a gravitational potential), Eq.~(\ref{eqn:pinch}) becomes \cite{Kolmes2018}
\begin{gather}
\big( n_a e^{\Phi_a/T} \big)^{1/Z_a} \propto \big( n_b e^{\Phi_b / T} \big)^{1/Z_b} , \label{eqn:generalizedPinch}
\end{gather}
where $T$ is the temperature of the system. 

Eqs.~(\ref{eqn:pinch}) and (\ref{eqn:generalizedPinch}) appear across a very wide range of systems and parameter regimes. These expressions, and special cases thereof, appear in the theory of classical transport \cite{Spitzer1952, Taylor1961ii, Braginskii1965, Bonnevier1966, Kolmes2018, HelanderSigmar, Ochs2018i, Ochs2018ii, Mlodik2020, Mitra2021, Kolmes2021HotIon}; in neoclassical transport including the Pfirsch-Schl\"uter, plateau, and banana regimes \cite{Connor1973, Hinton1974, Hirshman1981, Hirsch2008}; in the theory of plasma mass filters \cite{Krishnan1983, Geva1984, Prasad1987, Bittencourt1987, Grossman1991, Dolgolenko2017, Ochs2017}; in the study of non-neutral plasmas \cite{ONeil1981, Imajo1997, Dubin1999}; and, recently, in the theory of collisionally unmagnetized (low-Hall-parameter) cross-field transport \cite{Mlodik2021PartialMagnetization}. 
The ubiquity of this behavior, even in different systems whose dynamics are understood in terms of mutually inapplicable models, raises the question: can the appearance of this same behavior across all of these different systems be understood in terms of some kind of universal requirement? 

Ref.~\cite{Kolmes2020MaxEntropy} suggests that it can. If the maximum-entropy state is calculated subject to some fixed total energy and fixed particle populations, the result is the Boltzmann distribution. If an additional constraint fixes the net ion charge densities to some initial distribution, then the result is instead Eq.~(\ref{eqn:pinch}) or (\ref{eqn:generalizedPinch}), depending on the presence or absence of potentials $\Phi_s$. 
This constraint is physically motivated by the ambipolarity of cross-field collisional transport. 
This approach explains the behavior in Eqs.~(\ref{eqn:pinch}) and (\ref{eqn:generalizedPinch}) without having to specify any of the details of the dynamics (other than that the system must tend toward its maximum-entropy state and that it must respect the aforementioned constraints). 

However, the approach in Ref.~\cite{Kolmes2020MaxEntropy} has a significant drawback. 
Because that calculation approaches the problem in terms of the equilibrium thermodynamics of a closed system, it has no natural way to describe the effects of a temperature gradient. 
This is more important than it might sound. 
A properly oriented temperature gradient is known (both theoretically and experimentally) \cite{Rutherford1974, Wade2000, Dux2004, Hirsch2008, Helander2017, Newton2017} to mitigate and even reverse the accumulation of high-$Z$ ions described in Eq.~(\ref{eqn:pinch}), under the right circumstances. This ``temperature-screening" effect is one of the major strategies used by the operators of real confinement devices to prevent the (otherwise rather dire) accumulation of heavy impurities predicted by Eq.~(\ref{eqn:pinch}). 

The object of this paper is to demonstrate that the basic finding of Ref.~\cite{Kolmes2020MaxEntropy} -- that is, the universal role of an ambipolarity condition in giving rise to Eqs.~(\ref{eqn:pinch}) and (\ref{eqn:generalizedPinch}) -- still applies for systems that have temperature gradients.  
Moreover, this paper will show that the non-equilibrium theory can predict the temperature-screening effects observed in real systems, and that temperature screening depends on the details of the system's dynamics in a way that the results in Ref.~\cite{Kolmes2020MaxEntropy} did not. 

This paper is organized as follows. 
Section~\ref{sec:variational} describes the linear non-equilibrium formalism and coordinates to be used in the rest of the paper. 
Section~\ref{sec:ambipolarityDerivation} shows how an ambipolarity condition gives rise to Eq.~(\ref{eqn:pinch}), Eq.~(\ref{eqn:generalizedPinch}), and their generalizations with temperature screening. 
Section~\ref{sec:example} presents an example of how this formalism works in a simple slab system. 
Section~\ref{sec:discussion} is a discussion of these results. 
\ref{appendix:variational} makes a connection between this problem and non-equilibrium variational principles. 

\section{Thermodynamic Fluxes, Forces, and the Linear Regime} \label{sec:variational}

The transport in a non-equilibrium system can be described in terms of thermodynamic fluxes $J_i$ and forces $X_i$, such that the entropy production density can be written as 
\begin{gather}
\sigma = \sum_i J_i X_i. 
\end{gather}
The choices of $J_i$ and $X_i$ are not unique. 
In the literature on non-equilibrium thermodynamics, it is common to use 
\begin{gather}
\tilde{\mathbf{J}} = \begin{pmatrix} 
\mathbf{q} \\
m_1 n_1 \bv_1 \\
\vdots \\
m_N n_N \bv_N
\end{pmatrix}, 
\end{gather}
where $\mathbf{q}$ is the heat flux, $m_s$ is the mass of species $s$, $n_s$ is the number density of species $s$, and $\bv_s$ is the velocity of species $s$, and 
\begin{gather}
\tilde{\mathbf{X}} = \begin{pmatrix} 
\nabla(1/T) \\
- \nabla (\mu_1 / m_1 T) + \mathbf{F}_1 / m_1 T \\
\vdots \\
- \nabla (\mu_N / m_N T) + \mathbf{F}_N / m_N T
\end{pmatrix}, 
\end{gather}
where $T$ is the system temperature, $\mu_s$ is the internal chemical potential of species $s$, and $\mathbf{F}_s$ is any body force acting on species $s$ (we will assume for simplicity that all species have the same local temperature, though a version of this problem could be posed in which this was not the case). 

However, it is more convenient for present purposes to transform to a different set of forces and fluxes. This kind of transformation is described in Ref.~\cite{Boozer1992}. 
If $X_i = \sum_j M_{ij} \tilde{X}_j$ for some non-singular matrix of coefficients $M$ with inverse $M^{-1}$, then the appropriate fluxes $\mathbf{J}$ to associate with $\mathbf{X}$ are given by $J_i = \sum_j \tilde{J}_j (M^{-1})_{ji}$. 

With that in mind, let  
\begin{gather}
\mathbf{X} = \begin{pmatrix} 
\nabla(1/T) \\
- \nabla \log p_1 + \mathbf{F}_1 / T \\
\vdots \\
- \nabla \log p_N + \mathbf{F}_N / T
\end{pmatrix}, \label{eqn:generalizedForces}
\end{gather}
where $p_s \doteq n_s T$. The gradient of the chemical potential can be written as $\nabla (\mu_s / T) = \nabla \log ( p_s T^{-5/2} )$, so the appropriate transformation matrices are 
\begin{align}
M = \begin{pmatrix}
1 \\
-5 T / 2 & m_1 \\
-5 T / 2 & & m_2 \\
\vdots & & & \ddots \\
-5 T / 2 & & & & m_N
\end{pmatrix}
\end{align}
and 
\begin{align}
M^{-1} = \begin{pmatrix}
1 \\
5 T / 2 m_1 & m_1^{-1} \\
5 T / 2 m_2 & & m_2^{-1} \\
\vdots & & & \ddots \\
5 T / 2 m_N & & & & m_N^{-1}
\end{pmatrix}. 
\end{align}
Then the associated flux $\mathbf{J}$ is 
\begin{gather}
\mathbf{J} = \begin{pmatrix}
\mathbf{q} + (5/2) \sum_{s=1}^N p_s \bv_s \\
n_1 \bv_1 \\
\vdots \\
n_N \bv_N 
\end{pmatrix} . 
\end{gather}
Note that $\sigma = \tilde{\mathbf{J}} \cdot \tilde{\mathbf{X}} = \mathbf{J} \cdot \mathbf{X}$. 

For a system that is sufficiently close to equilibrium, $J_i$ and $X_i$ are linearly related, so that 
\begin{gather}
J_i = \sum_j L_{ij} X_j \label{eqn:linearity}
\end{gather}
for some matrix of coefficients $L_{ij}$. This is sometimes called the linear or Onsager regime, and the associated matrix $L$ is sometimes called the Onsager matrix. 
Much of the theory of plasma transport is concerned with detailed calculations of $L_{ij}$ for a particular system. 

Onsager's reciprocal relations state that the Onsager matrix is symmetric: that is, $L_{ij} = L_{ji}$. The original form of the theorem's proof does not apply in the presence of a magnetic field; in the presence of $\mathbf{B}$, the Onsager symmetry $L_{ij} = L_{ji}$ has traditionally been replaced by the Onsager-Casimir symmetry $L_{ij}(\mathbf{B}) = L_{ji}(-\mathbf{B})$ \cite{Onsager1931i, Onsager1931ii, Casimir1945}. However, recent results suggest that the Onsager symmetry may apply to cases with magnetic fields after all \cite{Bonella2014, Luo2020, Coretti2020, Carbone2022}, and in any event the coefficients in plasma systems are very often even functions of $\mathbf{B}$, in which case the distinction is moot \cite{Boozer1992}. 

\section{From Flux Constraints to Impurity Pinch} \label{sec:ambipolarityDerivation}

\ref{appendix:variational} describes how the principle of minimum entropy production leads to a set of constraints on the fluxes through the system. 
Of course, even in cases where Onsager symmetry does not hold, or in which Prigogine's minimum-entropy-production principle is otherwise inapplicable, it would be equally valid to consider a system in which \textit{all} fluxes are fixed by the system's boundary conditions. 
Either way, this section will consider the problem in which, for one reason or another, the fluxes $J_i$ are fixed, and especially the case where the $J_0$ may not vanish but the particle fluxes do vanish. 

For the calculation that follows, assume that Onsager symmetry does at least hold for the coefficients coupling the particle fluxes with thermodynamic forces associated with different species' pressure gradients (in other words, that $L_{ss'} = L_{s's}$ for any $s, s' \neq 0$; we do not require Onsager symmetry for the coefficients that couple to the temperature gradient and heat flux). This is a weaker assumption than is required for the minimum-entropy-production principle in \ref{appendix:variational}. 

Consider a one-dimensional system in a magnetic field (so that all spatial functions can be considered to depend only on some coordinate that measures position perpendicular to the field). 
Suppose the system obeys an ambipolarity constraint, much like the one discussed in Ref.~\cite{Kolmes2020MaxEntropy}. Such a constraint could be written as 
\begin{gather}
\sum_{s=1}^N Z_s J_s = 0. \label{eqn:protoAmbipolarity}
\end{gather}
Eq.~(\ref{eqn:protoAmbipolarity}) should be understood to apply at all times (not just in steady state); in other words, it is a constraint on the Onsager matrix. 
For Eq.~(\ref{eqn:protoAmbipolarity}) to hold for all possible thermodynamic forces, it must be true for any vector $\mathbf{Y}$ that 
\begin{gather}
\sum_{s=1}^N Z_s \sum_{i=0}^N L_{si} Y_i = 0. 
\end{gather}
This implies that 
\begin{gather}
\sum_{s=1}^N Z_s L_{sk} = 0 \hspace{15 pt} \forall k \in \{0, 1, \dots, N\}. \label{eqn:ambipolarity}
\end{gather}
Intuitively, this kind of constraint can be motivated in terms of the response of a particle to a force. If a particle of species $s$ is acted upon by a force $\mathbf{F}$, it will drift across the local magnetic field at a velocity $\mathbf{F} \times \bB / Z_s e B^2$. The $1/Z_s$ dependence in the cross-field motion means that interactions between particles tend to rearrange the particles in a way that satisfies Eq.~(\ref{eqn:protoAmbipolarity}). 
Eq.~(\ref{eqn:protoAmbipolarity}) is not, in general, an exact conservation law. 
Small uncompensated cross-field currents can appear for a wide variety of reasons (for example, inhomogeneities in the magnetic field). 
Ambipolarity is, however, typically a very good approximation for plasmas immersed in magnetic fields that are sufficiently strong and do not vary too quickly. 

Suppose the flux constraints are chosen so that the heat-flux term $J_0 \neq 0$, but so that the particle fluxes $J_s$ all vanish. 
Then for each species $s$, 
\begin{gather}
J_s = - \frac{L_{s0} T'}{T^2}  + \sum_{s'=1}^N L_{ss'} \bigg( \frac{F_{s'}}{T} - \frac{p_{s'}'}{p_{s'}} \bigg) = 0, \label{eqn:vanishingFlux}
\end{gather}
as per Eqs.~(\ref{eqn:generalizedForces}) and (\ref{eqn:linearity}). 
In the special case where the thermal coupling term $L_{s0} T'$ vanishes, this can be rewritten as 
\begin{gather}
\sum_{s'=1}^N Z_{s'} L_{ss'} \bigg( \frac{F_{s'}}{Z_{s'}} - \frac{p'_{s'}}{Z_{s'} n_{s'}} \bigg) = 0 . 
\end{gather}
Keeping in mind Eq.~(\ref{eqn:ambipolarity}), this is solved whenever 
\begin{gather}
\frac{F_{s'}}{Z_{s'}} - \frac{p_{s'}'}{Z_{s'} n_{s'}} = C
\end{gather}
for some constant $C$ that is the same for all $s'$. This can be integrated directly to get the condition that 
\begin{align}
&\bigg[ \frac{p_{a}}{p_{a0}} \exp \bigg( - \int_{x_0}^x \D x \, \frac{F_{a}}{T} \bigg) \bigg]^{1/Z_{a}} \nonumber \\
&\hspace{30 pt}= \bigg[ \frac{p_{b}}{p_{b0}} \exp \bigg( - \int_{x_0}^x \D x \, \frac{F_{b}}{T} \bigg) \bigg]^{1/Z_{b}} \label{eqn:pinchNoThermalForce}
\end{align}
for some integration constants $p_{a0}$ and $p_{b0}$ and reference point $x_0$. 
This reduces to Eq.~(\ref{eqn:generalizedPinch}) in the case where $T$ is constant, and to Eq.~(\ref{eqn:pinch}) in the case where $F_s = 0$. 
Note that this derivation had two key ingredients: the Onsager symmetry $L_{ss'} = L_{s's}$ and the ambipolarity condition. 
Also note that the Onsager coefficients themselves do not appear in Eq.~(\ref{eqn:pinchNoThermalForce}). 

The equilibrium conditions are further modified in the more general case where the thermal friction term $L_{s0} T'$ does not vanish. 
In the most general case, the condition cannot be written much more compactly than Eq.~(\ref{eqn:vanishingFlux}), though it more nearly resembles some of the forms seen in the literature if it is rewritten as 
\begin{align}
&- \frac{L_{s0} T'}{T^2} + \nonumber \\
&\sum_{s'=1}^N Z_{s'} L_{ss'} \bigg[ \frac{1}{Z_{s'}} \bigg( \frac{F_{s'}}{T} - \frac{p_{s'}'}{p_{s'}} \bigg) - \frac{1}{Z_s} \bigg( \frac{F_s}{T} - \frac{p_s'}{p_s} \bigg) \bigg] \nonumber \\
&\hspace{50 pt}= 0. \label{eqn:vanishingFluxDifference}
\end{align}
As is discussed in Ref.~\cite{Mlodik2021PartialMagnetization}, the equilibrium condition cannot always be written in the ``transitive" form seen in Eqs.~(\ref{eqn:pinch}), (\ref{eqn:generalizedPinch}), or (\ref{eqn:pinchNoThermalForce}), where the condition reduces to a self-consistent set of pairwise relations between all pairs of particle species. 

However, it is often useful to look at the special cases where it can be expressed in the transitive form. 
It can be written in this form in the absence of thermal forces; this is Eq.~(\ref{eqn:pinchNoThermalForce}). 
It can also be expressed in this form in the case where there are only two species. In that case, the zero-flux conditions on particle species $a$ and $b$ become 
\begin{align}
&- \frac{L_{a0} T'}{Z_b L_{ab} T^2} + \frac{1}{Z_{b}} \bigg( \frac{F_{b}}{T} - \frac{p_{b}'}{p_{b}} \bigg) - \frac{1}{Z_a} \bigg( \frac{F_a}{T} - \frac{p_a'}{p_a} \bigg) \nonumber \\
&\hspace{150 pt}= 0 \\
&- \frac{L_{b0} T'}{Z_a L_{ab} T^2} + \frac{1}{Z_{a}} \bigg( \frac{F_{a}}{T} - \frac{p_{a}'}{p_{a}} \bigg) - \frac{1}{Z_b} \bigg( \frac{F_b}{T} - \frac{p_b'}{p_b} \bigg) \nonumber \\
&\hspace{150 pt}= 0,
\end{align}
assuming $L_{ab} \neq 0$. These two conditions are the same, since in the two-species case Eq.~(\ref{eqn:ambipolarity}) implies $L_{a0} / Z_b = - L_{b0} / Z_a$. 
In this case, the condition can be integrated to get 
\begin{align}
&\bigg\{ \frac{p_{a}}{p_{a0}} \exp \bigg[ - \int_{x_0}^x \D x \, \frac{F_{a}}{T} \bigg] \bigg\}^{1/Z_{a}} \nonumber \\
&= \bigg\{ \frac{p_{b}}{p_{b0}} \exp \bigg[ - \int_{x_0}^x \D x \, \bigg( \frac{F_{b}}{T} - \frac{L_{a0} T'}{L_{ab} T^2} \bigg) \bigg] \bigg\}^{1/Z_{b}} . \label{eqn:twoSpeciesPinch}
\end{align}
In cases without thermal frictions, Eq.~(\ref{eqn:twoSpeciesPinch}) describes familiar behavior: precisely the peaking of high-$Z$ species found in Eqs.~(\ref{eqn:pinch}) and (\ref{eqn:generalizedPinch}). 
In cases with thermal frictions, the values of the Onsager coefficients begin to matter; depending on the dynamics of the particular system in question, temperature gradients may tend to flush high-$Z$ species from high-density regions, or they may tend to pull them in. 

Essentially the same formalism used in this section to describe temperature screening can also be used to understand systems with steady-state particle fluxes: the difference is simply that $J_s$ is allowed to be nonzero for $s \neq 0$. 
This problem has received less attention in the literature than temperature screening has, but there are scenarios in which it could be important; see Ref.~\cite{Mitra2021}. 
The analytic expressions for $n_s$ in Ref.~\cite{Mitra2021} can be recovered from Eqs.~(\ref{eqn:generalizedForces}) and (\ref{eqn:linearity}) by direct integration (or equivalently from Eq.~(\ref{eqn:vanishingFlux}) with a nonzero RHS). 
For cases with finite particle fluxes, the values of some Onsager coefficients appear in the equilibrium conditions. 

\section{A Simple Illustrative Example} \label{sec:example}

Consider, for example, classical collisional transport in a magnetized slab. Suppose the system has Cartesian coordinates with unit vectors $(\hat x, \hat y, \hat z)$, with a magnetic field $\mathbf{B} = B \hat z$ and all gradients in the $\hat x$ direction. Furthermore, suppose the plasma consists of electrons, hydrogen ions, and some heavy impurity ion species. Variables referring to the hydrogen will be denoted with the subscript $H$; variables related to the impurity will be denoted by the subscript $I$. Let the plasma be strongly magnetized, so that the cross-field transport of the electrons is slow enough to be ignored. 

Then the thermodynamic forces can be written as 
\begin{gather}
\mathbf{X} = \begin{pmatrix} 
\partial_x (1/T) \\
- \partial_x \log p_H \\
- \partial_x \log p_I 
\end{pmatrix}
\end{gather}
and the corresponding fluxes can be written as 
\begin{gather}
\mathbf{J} = \begin{pmatrix}
\mathbf{q} + (5/2) p_H v_{H,x} + (5/2) p_I v_{I,x} \\
n_H v_{H,x} \\
n_I v_{I,x} 
\end{pmatrix} . 
\end{gather}
Here $v_{s,x}$ is the $\hat{x}$-directed velocity of species $s$. Suppose, for simplicity, that the plasma is inviscid. 

Suppose boundary conditions fix some cross-field heat flux while requiring that the $\hat x$ particle fluxes vanish. 
The relevant components of the linear response matrix $L_{ij}$ can be calculated directly by considering the equations of motion. For the purposes of calculating $n_H$ and $n_I$, the thermal-conductivity-associated coefficient $L_{00}$ is unimportant (so long as the solution is expressed in terms of some self-consistent $T(x)$). Otherwise, in steady state, 
\begin{align}
m_H \bv_H \cdot \nabla \bv_H = e \bv_H \times \bB - \frac{\nabla p_H}{n_H} + \frac{\mathbf{R}_{HI}}{n_H} \\
m_I \bv_I \cdot \nabla \bv_I = Z_I e \bv_I \times \bB - \frac{\nabla p_I}{n_I} + \frac{\mathbf{R}_{IH}}{n_I} \, ,
\end{align}
where $e$ is the elementary charge and $\mathbf{R}_{ss'}$ is the friction force density between species $s$ and $s'$. This force will in general include both flow frictions and thermal frictions. In the case of a heavy impurity, it can be expressed \cite{HelanderSigmar} as 
\begin{gather}
\mathbf{R}_{HI} = m_H n_H \nu_{HI} \bigg[ \bv_I - \bv_H + \frac{3 T'}{2 e B} \, \hat y \bigg] \\
\mathbf{R}_{IH} = m_I n_I \nu_{IH} \bigg[ \bv_H - \bv_I - \frac{3 T'}{2 e B} \, \hat y \bigg]. 
\end{gather}
Here $\nu_{ss'}$ is the collision frequency between species $s$ and $s'$, and the conservation of momentum requires that $m_H n_H \nu_{HI} = m_I n_I \nu_{IH}$. 

Dropping the advective $\bv_s \cdot \nabla \bv_s$ terms on the left-hand side of the equations of motion (since these are quadratic in $\mathbf{J}$), and dropping the $\hat x$ component of the flow frictions (since these ultimately contribute to $\mathbf{J}$ at a higher order in $m_s \nu_{ss'} / Z_s e B$, which is a small parameter in a strongly magnetized plasma), the equations of motion can be rewritten as 
\begin{align}
\mathbf{J} = \frac{m_H n_H \nu_{HI} T}{e^2 B^2} \begin{pmatrix}
\ell_{00} & \ell_{01} & \ell_{02} \\
3 T / 2 & 1 & - 1 / Z_I \\
- 3 T / 2 Z_I & - 1 / Z_I & 1 / Z_I^2
\end{pmatrix} \mathbf{X} . \label{eqn:particularL}
\end{align}
Here $\ell_{00}$, $\ell_{01}$, and $\ell_{02}$ are arbitrary matrix entries. 
They would be specified by a temperature evolution equation, but they are not necessary here. 
Note that Eq.~(\ref{eqn:particularL}) is consistent with Eq.~(\ref{eqn:ambipolarity}). 
In this system, Eq.~(\ref{eqn:twoSpeciesPinch}) becomes 
\begin{align}
\frac{p_I}{p_{I0}} = \bigg( \frac{p_H}{p_{H0}} \bigg)^{Z_I} \bigg( \frac{T}{T_0} \bigg)^{-3 Z_I/2} . 
\end{align}
This is one of the simplest examples that can exhibit temperature screening of impurities. 
In other regimes, peaked temperature profiles can have the opposite effect (pulling high-$Z$ impurities into the high-temperature regions rather than pushing them away), according to differences in $L_{s0}$. 

\section{Discussion} \label{sec:discussion}

In the absence of thermal frictions, in the linear (Onsager) non-equilibrium regime, this paper has shown that the relative cross-field accumulation of different ion species represented in results like Eqs.~(\ref{eqn:pinch}) and (\ref{eqn:generalizedPinch}) follows from two key conditions: (1) an ambipolarity condition on the flows of different species and (2) symmetry of the Onsager coefficients coupling the flow of one particle species to a thermodynamic force acting on another species. 
Apart from these conditions, no other details of the system's dynamics need to be specified. 
This extends the argument from Ref.~\cite{Kolmes2020MaxEntropy} to provide a unified explanation for the accumulation of high-$Z$ species in cases with temperature gradients but without thermal frictions. 

In cases with thermal frictions, this paper has shown that temperature screening effects arise naturally from the same formalism, and can be related in a simple way to the Onsager coefficients (though of course, temperature screening has been derived for many particular systems using the Onsager formalism before). 
This dependence on the Onsager coefficients is a significant difference: temperature screening can still be explained in terms of a generic linear-response theory, but in order to calculate the resulting equilibria it is necessary to compute some of the $L_{ij}$ (as opposed to the case without thermal frictions, where nothing need be known about the Onsager coefficients except that they enforce ambipolarity). 

The distinction between cases with and without significant thermal frictions is formally a question of the relative sizes of $L_{s0} T' / T^2$ and the other terms in Eq.~(\ref{eqn:ambipolarity}). 
It is clear that thermal frictions can be neglected when the temperature gradient vanishes or is sufficiently small. 
However, in general, how small that gradient has to be will depend on the details of the dynamics of the particular system in question (that is, on the actual value of $L_{s0}$). 
For example, in the simple case discussed in Section~\ref{sec:example}, thermal frictions can be considered significant whenever $T' / T$ is at least comparable to $n_H' / n_H$ or $n_I' / n_I$. 

The emphasis on $\nabla T$-dependent effects was motivated by the prominence of these effects in the field, both theoretically and in experimental studies \cite{Rutherford1974, Wade2000, Dux2004, Hirsch2008, Helander2017, Newton2017}. 
However, an essentially identical analysis could be used for cases in which the thermodynamic force vector $\mathbf{X}$ included forces other than the temperature and pressure gradient terms described in Eq.~(\ref{eqn:generalizedForces}). 
There are a number of situations in which additional thermodynamic forces may be important; see, for example, Refs.~\cite{Boozer1992, Jou, Sendra2021, Saluto2022}. 

The discussion in this paper treats the electrons as stationary. 
In collisional cross-field transport, this is typically a reasonable assumption. 
The smallness of the electron gyroradii means that classical processes move them across field lines on a timescale that is slow compared to ion-ion transport. 
Indeed, this same assumption is at least implicitly present in other derivations of Eqs.~(\ref{eqn:pinch}) and (\ref{eqn:generalizedPinch}). 
However, if for a given system the electron transport was not slow, there is no reason why electrons could not be included in the transport matrices as a species with charge $Z_e = -1$. 
Results like Eqs.~(\ref{eqn:pinch}) and (\ref{eqn:generalizedPinch}) have very different implications if they also apply to the electron population. 
For example, if applied to all electron and ion species in a quasineutral plasma, Eq.~(\ref{eqn:pinch}) implies that the density profiles must be flat. 
This makes sense, since on the longer timescales over which electrons can cross field lines collisionally, the plasma typically escapes from magnetic confinement. 

The focus here has been on cross-field dynamics. 
Of course, a related set of issues are important in unmagnetized plasma systems \cite{Amendt2010, Kagan2012, Kagan2014, Kagan2014ii, Zhang2020}, but these systems do not generally have the same ambipolarity constraints that appear in magnetized systems. 

The present investigation is confined to the comparatively settled areas of near-equilibrium thermodynamics. 
Much of the literature on plasma transport is concerned with this regime. 
The further reaches of non-equilibrium thermodynamics, particularly for systems far from equilibrium, would require a different theory. 
It is not necessarily clear that we should expect results like Eq.~(\ref{eqn:pinch}) and (\ref{eqn:generalizedPinch}) to continue to apply outside of the near-equilibrium regime.

\section*{Acknowledgements}
\noindent The authors thank Per Helander, whose comments prompted this investigation. 
This work was supported by Cornell NNSA 83228-10966 [Prime No. DOE (NNSA) DE-NA0003764] and by NSF PHY-1805316.

%

\appendix
\section{Flux Constraints from a Variational Principle} \label{appendix:variational}

The argument in Ref.~\cite{Kolmes2020MaxEntropy} was based on calculating the maximum-entropy state in an equilibrium system. 
There is no single universally accepted variational principle that plays an analogous role in non-equilibrium systems. 
In fact, the use of variational principles in non-equilibrium thermodynamics is an active area of research, and a number of authors have undertaken to develop generally applicable variational principles for these systems \cite{Dewar2003, Dewar2005, Niven2009, Niven2010}. 

In cases where Onsager symmetry holds, one non-equilibrium variational principle is the principle of minimum entropy production \cite{Klein1954, Prigogine, DeGroot}, which is valid in systems close to equilibrium. This variational principle is not strictly necessary to the results in the rest of this paper, and it comes with a number of serious limitations \cite{Landauer1975}, but it does establish a useful parallel with the maximum-entropy principle in Ref.~\cite{Kolmes2020MaxEntropy}, so we will briefly discuss it here, roughly following the discussion in Ref.~\cite{DeGroot}. 

In the Onsager regime, the total entropy production over some volume $\mathcal{V}$ is given by 
\begin{gather}
\dot S = \int_\mathcal{V} \D^3 \br \, \sum_{i=0}^N J_i X_i = \int_\mathcal{V} \D^3 \br \, \sum_{i=0}^N \sum_{j=0}^N L_{ij} X_i X_j .
\end{gather}
In the absence of any further constraints, $\dot S$ is minimized when 
\begin{gather}
\sum_{j=0}^N (L_{ij} + L_{ji}) X_j = 0. 
\end{gather}
In cases were Onsager symmetry holds, this becomes 
\begin{gather}
J_i = 0. 
\end{gather}
In other words, the entropy production vanishes when the fluxes vanish. 
In many cases of interest, there is some additional constraint on some of the fluxes or forces. If, for some $k$, $J_k$ is fixed by a constraint, then this simply becomes 
\begin{gather}
J_{i \neq k} = 0. 
\end{gather}
That is, any unconstrained fluxes vanish.

\bibliographystyle{apsrev4-1} 
\bibliography{../../../Dropbox/Master.bib}

\end{document}